\documentclass{article}
\usepackage[super]{cite}
\usepackage{graphicx}

\usepackage{braket}
\usepackage{amsmath,amssymb}
\usepackage{amsfonts}
\usepackage{graphicx}
\usepackage{comment}
\usepackage{graphicx}
\usepackage{epsfig}
\usepackage{dcolumn}
\usepackage[up]{subfigure}
\usepackage{adjustbox}
\usepackage{float}
\usepackage{dsfont}	
\usepackage{relsize}
\usepackage{verbatim}
\usepackage{float}
\usepackage{ragged2e}
\usepackage{braket}
\usepackage{color}
\usepackage{xcolor}
\usepackage[ruled,vlined]{algorithm2e}

\SetCommentSty{mycommfont}

\SetKwInput{KwInput}{Input}                
\SetKwInput{KwOutput}{Output} 
\SetKw{Continue}{continue}
\SetKw{Break}{break}

\usepackage{arxiv}

\usepackage[utf8]{inputenc} 
\usepackage[T1]{fontenc}    
\usepackage{hyperref}       
\usepackage{url}            
\usepackage{booktabs}       
\usepackage{amsfonts}       
\usepackage{nicefrac}       
\usepackage{microtype}      
\usepackage{lipsum}
\usepackage{graphicx}
\graphicspath{ {./images/} }

\title{Faster Search of Clustered Marked States with Lackadaisical Quantum Walks}

\author{Amit Saha$^{1,3*}$, Ritajit Majumdar$^2$, Debasri Saha$^{1}$, Amlan Chakrabarti$^{1}$, Susmita Sur-Kolay$^2$\\~\\

$^1$A. K. Choudhury School of Information Technology, University of Calcutta\\
$^2$Advanced Computing \& Microelectronics Unit, Indian Statistical Institute\\
$^3$Atos, Pune, India\\~\\
$^*$abamitsaha@gmail.com}

\date{}

\begin{document}
\maketitle
\begin{abstract}
The nature of discrete-time quantum walk in the presence of multiple marked states has been studied by Nahimovs and Rivosh. They introduced an exceptional configuration of clustered marked states $i.e.,$ if the marked states are arranged in a $\sqrt{k} \times \sqrt{k}$ cluster within a $\sqrt{N} \times \sqrt{N}$ grid, where $k=n^{2}$ and $n$ an odd integer. They showed that finding a single marked state among the multiple ones using quantum walk with AKR (Ambainis, Kempe and Rivosh) coin requires $\Omega(\sqrt{N} - \sqrt{k})$ time. Furthermore, Nahimov and Rivosh also showed that the Grover's coin can find the same configuration of marked state both faster and with higher probability compared to that with the AKR coin. In this article, we show that using lackadaisical quantum walk, a variant of a three-state discrete-time quantum walk on a line, the success probability of finding all the clustered marked states of this exceptional configuration is nearly 1 with smaller run-time.  We also show that the weights of the self-loop suggested for multiple marked states in the state-of-the-art works are not optimal for this exceptional configuration of clustered mark states. We propose a range of weights of the self-loop from which only one can give the desired result for this configuration. 
\end{abstract}


\section{Introduction}

As development of quantum computers has progressed  significantly in the last decade, quantum algorithms for new problems, which provide potential speedup over their classical counterparts \cite{nielsen_chuang_2010} are also being studied enthusiastically. Grover's algorithm \cite{grover1996fast} is one of the quantum algorithms that has quadratic speedup for searching a marked location in an unsorted database. Quantum walk \cite{PhysRevA.48.1687} is another such quantum algorithm that can be a prospective candidate for solving search problems \cite{Shenvi_2003, ambainis2004quantum, aharonov2000quantum, magniez2003quantum} with considerable speedup over their classical versions. Quantum walks (QW) have two main variants, namely continuous-time (CTQW) \cite{Childs_2003} and discrete-time \cite{ambainis2003quantum, Tregenna_2003} (DTQW). The probability distribution of the particles for both the variants of the quantum walk, soars quadratically faster in position space compared to the classical random walk \cite{Kempe03, Venegas_Andraca_2012, Kendon_2006}. A CTQW evolves under a Hamiltonian which is defined with respect to a graph, and no coin operator is required. A DTQW evolves through a quantum coin operator. In DTQW on a 1D grid (line), a two-state quantum coin operator has been used.

\begin{figure}[ht!]
    \centering
    \includegraphics[scale=.5]{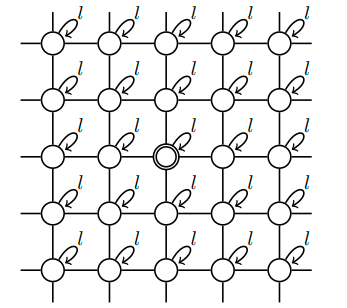}
    \caption{A 2D grid of $N = 5 \times 5$ vertices with a self-loop of weight $l$ at each vertex. The
boundaries are periodic. A marked vertex is indicated by a double circle. \cite{Wong_2016}}
    \label{fig:l}
\end{figure}

The concept of lazy quantum walk \cite{Childs_2009} (LQW), where the walker has equal probability of staying put, was later introduced incorporating a three-state quantum coin operator on a line. This helps to establish a relationship between CTQW and DTQW \cite{Childs_2009}. A small variation in lazy quantum walk gave birth to a breakthrough algorithm $i.e.,$ lackadaisical quantum walk \cite{Wong_2015}, which gives algorithmic speed up over the previous ones \cite{Wong_2018, hyer2020analysis, rhodes2020search, Rhodes_2019, Wang_2017, wang2017adjustable}. Lackadaisical quantum walk assigns a self-loop of weight $l$ to each
vertex (which can be varied as necessary; refer Figure \ref{fig:l}), so that
the walker has non-zero probability of staying put.

Wong \cite{Wong_2016} showed that the lackadaisical quantum walk, having self-loop of weight $4/N$ for each vertex, searches a single marked state with a $\sim 1$ success probability in
$O(\sqrt{N log N})$ steps, thus yielding an $O(\sqrt{log N})$ improvement over the DTQW. Later on , the authors \cite{nahimovs2018lackadaisical, saha2018search, Giri_2019, nahimovs2021lackadaisical, decarvalho2021applying} have studied the advantage of lackadaisical quantum walk over DTQW for multiple marked states. But, these works have failed to provide a generalized solution for the exceptional configuration of multiple marked states arranged in a $\sqrt{k} \times \sqrt{k}$ cluster within a $\sqrt{N} \times \sqrt{N}$ grid using lackadaisical quantum walk over DTQW.

In this paper, we propose a weight $l=\dfrac{4}{N*(k+1)} \pm \delta$ for the self-loop through numerical simulation, where $\delta$ is chosen such that
$\dfrac{4}{N*(k+2)} \le l \le \dfrac{4}{N*k}$. 
Out of this range of weights, only one weight can attain the success probability $\sim 1$ for finding all the clustered marked states with run-time lower than the state-of-the-art algorithms because no amplitude amplification is required.

This article is organized as follows: we briefly discuss  standard quantum walk and lackadaisical quantum walk in a two-dimensional grid in Section 2. We discuss the works  related to our proposed work in Section 3. In Section 4, we present our proposed methodology for cluster of marked states on a two-dimension grid using lackadaisical quantum walk. Finally, we  conclude in Section 5.

\section{Background}
A few preliminaries of quantum random walk are presented now.

\subsection{Discrete time quantum walk in 2-Dimensions}
A quantum random walk consists of a position Hilbert space $H_p$ and  a coin Hilbert space $H_c$. A quantum state consists of these two degrees of freedom, $\ket{c} \otimes \ket{v}$ where $\ket{c} \in H_c$ and $\ket{v} \in H_p$. A step in quantum walk is a unitary evolution $U = S.(C \otimes I)$ where $S$ is the shift operator and $C$ is the coin operator, which acts only on the coin Hilbert space $H_c$. If we consider a $\sqrt{N} \times \sqrt{N}$ grid, then the quantum walk starts in a superposition of states given by
\begin{equation}
    \ket{\psi(0)} = \frac{1}{\sqrt{4N}}(\sum_{i=1}^{4}\ket{i} \otimes \sum_{x,y=1}^{\sqrt{N}}\ket{x,y})
\end{equation}

where (i) each location $(x,y)$ corresponds to a quantum register $\ket{x,y}$ with $x,y \in \{1, 2, \hdots, \sqrt{N}\}$ and (ii) the coin register $\ket{i}$ with $i \in \{\leftarrow, \rightarrow, \uparrow, \downarrow\}$. The most often used transformation on the coin register is the Grover's Diffusion transformation $D$

\begin{equation}
\label{eq:diffusion}
    D = \frac{1}{2}\begin{pmatrix}
    -1 & 1 & 1 & 1\\
    1 & -1 & 1 & 1\\
    1 & 1 & -1 & 1\\
    1 & 1 & 1 & -1
    \end{pmatrix}
\end{equation}

The Diffusion operator can also be written as $D = 2\ket{s_D}\bra{s_D} - I_4$, where $\ket{s_D} = \frac{1}{\sqrt{4}}\sum_{i=1}^{4}\ket{i}$.\\

The transformation creates a superposition of the coin states $\ket{i}$, which in turn governs the shift operation. Multiple shift operators have been proposed in the literature \cite{ambainis2011search} out of which, in this paper, we have used the Flip-Flop shift transformation $S$ \cite{ambainis2004coins} whose action on the basis states are as follows:
\begin{align}
    \ket{i,j,\uparrow} & = \ket{i,j-1,\downarrow}\\
    \ket{i,j,\downarrow} & = \ket{i,j+1,\uparrow}\\
    \ket{i,j,\leftarrow} & = \ket{i-1,j,\rightarrow}\\
    \ket{i,j,\rightarrow} & = \ket{i+1,j,\leftarrow}.
 \end{align}

From Equations 3-6, we can infer that the Flip-Flop shift transformation changes the value
of the direction register to the opposite after moving to an adjacent position state. 

It is easy to see that $\ket{\psi(t)}$ is a $+1$ eigenstate of the operator $U = S.(D \otimes I)$ for any $t \in \mathbb{Z}$. A perturbation is created in the quantum state by applying the coin operator $-I$ instead of $D$ for marked locations. A general quantum walk algorithm applies this unitary operation (appropriately for the marked and the unmarked states) $t$ times to create the state $\ket{\psi(t)}$ such that $\braket{\psi(t)|\psi(0)}$ becomes close to 0. Measurement of the state $\ket{\psi(t)}$ is expected to give the marked location with high probability.

\subsection{Lackadaisical quantum walk in 2-Dimensions}

In a 2-dimensional lackadaisical quantum walk, the degree of freedom of the coin is five-dimensional, i.e. $i \in \{\leftarrow, \rightarrow, \uparrow, \downarrow, . \}$. The flip-flop transformation conditioned on the $\ket{.}$ coin state is
\begin{equation}
    S(\ket{i,j} \otimes \ket{.}) = \ket{i,j} \otimes \ket{.}
\end{equation}

If $l$ self-loops are allowed, then the Coin operator is $D = 2\ket{s_D}\bra{s_D} - I_5$, where 

\begin{equation}
    \ket{s_D} = \frac{1}{\sqrt{4+l}}(\ket{\uparrow} + \ket{\downarrow} + \ket{\leftarrow} + \ket{\rightarrow} + \sqrt{l}\ket{.})
\end{equation}

\subsection{Clustered marked states in a 2-dimensional grid}

\begin{figure}[ht!]
    \centering
    \includegraphics[scale=.5]{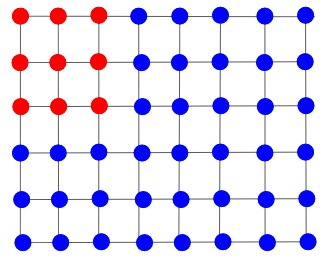}
    \caption{$\sqrt{k} \times \sqrt{k}$ marked locations (red nodes) in a $\sqrt{N} \times \sqrt{N}$ grid}
    \label{fig:cluster}
\end{figure}

The notion of clustered marked states was introduced by Nahimovs et al. \cite{nahimovs2015exceptional} In accordance with that paper, we consider the walk on a $\sqrt{N} \times \sqrt{N}$ grid where there are $k$ marked locations arranged in a $\sqrt{k} \times \sqrt{k}$ cluster (Fig.~\ref{fig:cluster}).

\section{Related Works}


After the success of Grover's algorithm as a quantum search algorithm, Benioff \cite{benioff2000space} first showed that if the $N$ data points are arranged in a $\sqrt{N} \times \sqrt{N}$ grid, then the quantum speedup is lost. Since then, research has been carried out to design faster algorithms to search an unsorted database arranged in a two or higher dimensional grid. Ambainis et al. \cite{inproceedings} proposed an algorithm (referred henceforth as the AKR algorithm) based on quantum random walk with AKR coin. The diffusion operators of the AKR coin for the marked and unmarked states are $-I$ and $D$ respectively, where $D$ is as in Eq.~(\ref{eq:diffusion}). This algorithm can detect a marked state with probability $\mathcal{O}(\frac{1}{logN})$ in $\mathcal{O}(\sqrt{NlogN})$ time. In order to increase the success probability, amplitude amplification is necessary, which has a time complexity of $\mathcal{O}(\sqrt{logN})$. This gives an overall running time of the algorithm to be $\mathcal{O}(\sqrt{N}logN)$. Childs and Goldstone  \cite{Childs_2004, Wong_2016} matched this runtime with a CTQW. Ambainis et al. \cite{ambainis2011search} henceforth, proposed another algorithm, which does not require amplitude amplification, and can perform the search in $\mathcal{O}(\sqrt{NlogN})$ time. Further works have been carried out to study quantum walk algorithms in other graph structures, but in this article we consider only the two-dimensional grid. 

Most of the quantum walk based search algorithms consider one or two marked locations.  Nahimovs and Rivosh \cite{nahimovs2015quantum} considered searching for multiple marked states within a $\sqrt{N} \times \sqrt{N}$ grid. The authors showed that if $k$ marked states are clustered in a $\sqrt{k} \times \sqrt{k}$ block, then the algorithm of \cite{ambainis2004coins} can perform the search in $\Omega(\sqrt{N} - \sqrt{k})$ time by using a Grover's coin, where the diffusion operators for the marked and unmarked states are $D$ and $-D$ respectively. Whereas if the $k$ marked locations are distributed uniformly over the grid, then the algorithm requires $\mathcal{O}(\sqrt{\frac{N}{k}log\frac{N}{k}})$ time. The authors \cite{nahimovs2015exceptional} also showed that if $k$ marked states are grouped in a $\sqrt{k} \times \sqrt{k}$ block, where $k=n^{2}$ and $n$ is an even number $\ge 2$, then the quantum walk always fails to find any of the marked locations.

In \cite{saha2018search}, the authors extended the model of \cite{Wong_2015} to multiple marked states for the first time, arranged in a $\sqrt{k} \times \sqrt{k}$ cluster, where $k=n^{2}$ and $n$ is an odd number, and showed by simulation that adjusting the weight of the self-loop to $\frac{4}{N(k + \lfloor{\frac{\sqrt{{k}}}{2}}\rfloor)}$, the probability of finding the marked states increases by $\sim 0.2$. The number of steps required is less than that of the quantum walk with no self loop or Grover's coin. The authors also showed that adjusting the weight of the self loop in \cite{Wong_2015} does not work for multiple mark states. But, the limitation of \cite{saha2018search} was that the authors used only $k=9$ and $N \le 30$. Later on, Nahimovs \cite{nahimovs2018lackadaisical, nahimovs2021lackadaisical} and Giri et al. \cite{Giri_2019} provided an adjustable weight of the self loop for multiple marked states, which are $l=\dfrac{4(k-\sqrt{k})}{N}$ \cite{nahimovs2018lackadaisical}, $l=\dfrac{4*k}{N}$ \cite{Giri_2019} respectively for different grid size $N$. Albeit, their weights also fail to search a cluster of marked states. In this paper, we provide a generalized solution for this exceptional configuration of clustered marked states, which is thoroughly discussed in the next section.

\section{Lackadaisical quantum walk for clustered marked states in a 2-dimensional grid}

\begin{figure}[ht!]
    \centering
    \includegraphics[scale=0.5]{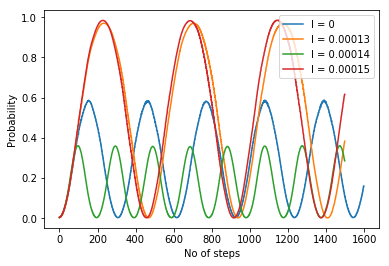}
    \caption{Success probability as a function of time for various $l$ in a 
2D grid of size $50 \times 50$.}
    \label{fig:convergence_weight}
\end{figure}

 As discussed above, the authors \cite{nahimovs2015exceptional} have used quantum walk where the weight on the self loop is $0$, or in other words, is not lackadaisical. They have used Grover's coin in the AKR algorithm and exhibited that Grover's coin significantly outperforms the AKR coin with respect to success probability of finding clustered marked states in a 2-dimensional grid. In \cite{Wong_2015}, the author has shown that using a weight of $\frac{4}{N}$ for the self loop, the probability of finding a single marked state increases with respect to that for non-lackadaisical walk. However, for multiple marked states, this weight provides a probability poorer than non-lackadaisical walk\cite{saha2018search, nahimovs2018lackadaisical, Giri_2019}. In this paper, we have proposed a weight of the self loop, $l=\dfrac{4}{N*(k+1)} \pm \delta$, such that $\dfrac{4}{N*(k+2)} \le l \le \dfrac{4}{N*k}$, for finding clustered mark states which leads to a total success probability of nearly $1$ in steps fewer than the state-of-the-art models of quantum walk for different grid sizes, where the probability of finding each marked state is \emph{(total probability)/$k$}. The running time and the success probability of the lackadaisical quantum walk completely depends on a weight of the
self-loop $l$. Figure \ref{fig:convergence_weight} shows the evolution of the probability of finding the clustered mark vertices of $k=9$ for the lackadaisical quantum walk on a grid of size $N = 50 \times 50$ for various values of $l$ through simulation. As one can see different values of $l$ result in different success
probabilities and number of steps till the first peak. When $l = 0$, i.e., the non-lackadaisical
quantum walk, the success probability reaches $\sim 0.6$. The success probability becomes $\sim 1$ when $l \approx 0.00015$, which is approximately $\dfrac{4}{N*(k+1)}$. We also observe that with the increase in steps, the success probability is  periodic and the value of the peaks remain the same through out, which implies that the first peak is the optimum value. The algorithmic presentation of our proposed work is briefly illustrated in Algorithm \ref{algo}. 

\begin{algorithm}
\SetAlgoLined  
\KwInput{ $N$, $k$} \vspace{.2cm}
\tcp{$N$, the size of the 2-dimensional grid; $k$, the number of clustered mark states, which can be described as a set of clustered mark vertices $K(x,y)$ correspond to a quantum register $\ket{x,y}$ with $x,y \in \{1, \hdots, \sqrt{N}\}$.} \vspace{.2cm}
\KwOutput{$t$, $prob$} \vspace{.2cm}
\tcp{$t$, the number of lackadaisical quantum walk steps taken to find the clustered mark states with $prob$, the success probability  nearly 1.}

\vspace{.2cm}
\tcp{Define $l$, the weight of each self loop }
 Choose a value of $\delta$ such that $l = \dfrac{4}{N*(k+1)} \pm \delta$ satisfies $\dfrac{4}{N*(k+2)} \le l \le \dfrac{4}{N*k}$.\\
\tcp{Define $\ket{\psi(0)}$, the initial quantum state}
\vspace{.2cm}
 $\ket{\psi(0)} = \frac{1}{\sqrt{N(4+l)}}(\sum_{i=1}^{5}\ket{i} \otimes \sum_{x,y=1}^{\sqrt{N}}\ket{x,y})$;\\ \vspace{.2cm}

\vspace{.2cm}

\tcp{Define $\ket{i}$, the coin register with $i \in \{\leftarrow, \rightarrow, \uparrow, \downarrow, .\}$, as the lackadaisical coin operator $D$}
\vspace{.2cm}
 $D=2\ket{s_D}\bra{s_D} - I_5$;\\ \vspace{.2cm}
\tcp{$\ket{s_D}=\frac{1}{\sqrt{4+l}}(\ket{\uparrow} + \ket{\downarrow} + \ket{\leftarrow} + \ket{\rightarrow} + \sqrt{l}\ket{.})$.}
\vspace{.2cm}

\tcp{Define $U$, the walk operator }\vspace{.1cm}
$U = S.(D \otimes I)$;\\
\tcp{$S$ is Flip-Flop Shift transformation:} 
\textcolor{blue}{{\footnotesize $\ket{i,j,\uparrow}  = \ket{i,j-1,\downarrow}$,\\
   $\ket{i,j,\downarrow}  = \ket{i,j+1,\uparrow}$,\\
    $\ket{i,j,\leftarrow}  = \ket{i-1,j,\rightarrow}$,\\
    $\ket{i,j,\rightarrow} = \ket{i+1,j,\leftarrow}$,\\
    $\ket{i,j, .} = \ket{i,j, .}$.}}\\
\tcp{Define the clustered marked vertices}
\vspace{.1cm}
$K(x, y)=-K(x, y)$, $\forall K(x, y)$;\\
\vspace{.1cm}
\tcp{Initialize the success probability and the number of steps} \vspace{.1cm}
\emph{prob}=0,
\emph{t}=0;
\vspace{.1cm}

\While{$t \ge 0$}{
  \tcp{The evolution of lackadaisical quantum walk}
  $\ket{\psi(t+1)}=U\ket{\psi(t)}$\;
  \eIf{$prob > \sum\limits_{\ket{x,y} \in K(x, y)}\lvert\braket{\psi|x, y}\rvert^2$}{
   \tcp{If first peak is found}
   break\;
   }{
   \tcp{increment the number of steps}
   $t=t+1$\;
   \tcp{Save the current success probability for the next step}
   $prob=\sum\limits_{\ket{x,y} \in K(x, y)}\lvert\braket{\psi|x, y}\rvert^2$\;
  }
 }
\KwRet{$t$, $prob$}
\caption{LQW for clustered marked states in a 2-dimensional grid}
 \label{algo}
\end{algorithm}

\begin{figure}[ht!]
    \centering
    \includegraphics[scale=0.45]{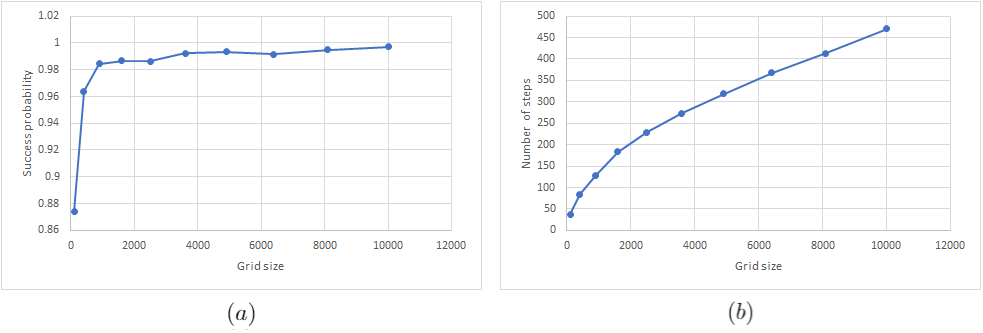}
    \caption{LQW  with  $l=\dfrac{4}{N*(k+1)} \pm \delta$, where $\dfrac{4}{N*(k+2)} \le l \le \dfrac{4}{N*k}$for $k = 9$  and different values of grid size $N$: ($a$) Success probability; and ($b$) Number of steps of LQW }
    \label{fig:proposed}
\end{figure}

Our simulations exhibit that using the weight of the self loop $l=\dfrac{4}{N*(k+1)} \pm \delta$, the success probability is close to $1$ for different grid sizes ranging from $10 \times 10$ to $100 \times 100$, which is shown in Table \ref{tbl:k=9}. We also observe that $\delta$ varies between $-0.00001$ to $0.0004$ for the aforesaid range of grid size. In Figure~\ref{fig:proposed}, we represent the results of our simulations graphically. 

\begin{table}[ht!]
\centering
\caption{Success probability and Number of steps for proposed LQW with $l=\dfrac{4}{N*(k+1)} \pm \delta$ for $k = 9$ and different values of grid size $N$}
{\begin{tabular}{@{}cccccc@{}} \hline
Grid size & $\dfrac{4}{N*(k+1)}$ & $\delta$ & Weight ($l$) & success probability  & Steps  \\
\hline
100   & 0.004 & 0.0004    & 0.0044   &  0.874064    & 37          \\ 
400    & 0.001 &  0    &  0.001 & 0.963963    & 84         \\ 
900    &  0.00044 & 0.00005     & 0.000490   & 0.984277    & 128        \\ 
1600      &  0.00025 &  0  & 0.00025   & 0.986564    & 183       \\ 
2500    & 0.00016 &   -0.00001   & 0.00015  & 0.986460    & 229           \\ 
3600    & 0.000111 &   0   & 0.000111  &  0.992471   & 273     \\ 
4900     & 0.000081 &  0.000001   & 0.000082   & 0.993541    &  319       \\ 
6400    & 0.000062 &  -0.000002    & 0.000060  & 0.991610  & 368      \\ 
8100   & 0.000049 &  0     & 0.000049  & 0.994785    & 413      \\ 
10000   & 0.000040 &   0.000003   & 0.000043  & 0.997134    & 471       \\ 
\hline
\end{tabular}
\label{tbl:k=9}}
\end{table}

\begin{figure}[ht!]
    \centering
    \includegraphics[scale=0.45]{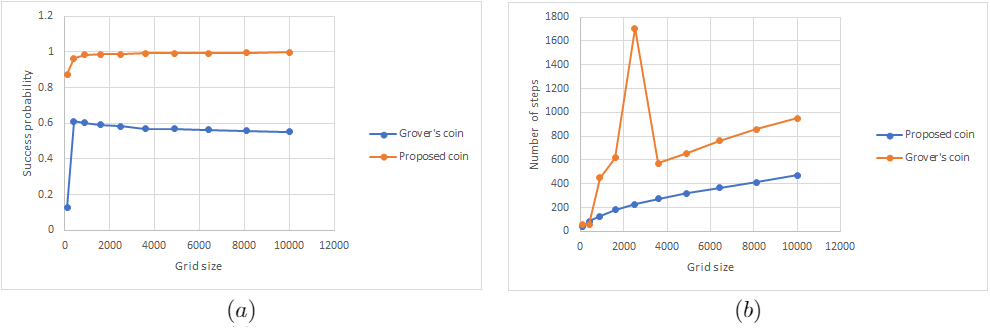}
    \caption{Comparison of the LQW algorithm with Grover's coin vs. proposed coin for $k = 9$ and different values of grid size $N$: ($a$) Success probability; ($b$) Number of steps.}
    \label{fig:compgrov}
\end{figure}

In Table \ref{Proposed_vs_Grover_k=9}, we have presented a comparative study of the Grover's coin and our proposed approach.  We have carried out simulations for $k = 9$, i.e., 9 marked states arranged in a $3 \times 3$ cluster for different grid sizes ranging from $10 \times 10$ to $100 \times 100$. The results clearly show that our proposed approach of lackadaisical quantum walk outperforms Grover's coin based quantum walk approach both in terms of success probability and run-time, which is shown in Figure \ref{fig:compgrov}. Our lackadaisical quantum walk approach finds all the marked states with success probability nearly 1 with the number of steps linear in the size of the grid, which may be considered as  optimal. For the Grover's coin based quantum walk, the algorithm needs to be iterated $\sqrt{\log{N}}$ times more in order to attain the success probability   $\sim 1$,.

\begin{table}[ht!]
\centering

\caption{Number of steps and success probability of the LQW algorithm with the Grover's coin vs. proposed coin, for $k = 9$ in different values of grid size $N$.}
{\begin{tabular}{@{}|c|cc|cc|@{}} \hline
 & Grover's Coin & & Proposed Coin & \\
 
\hline
Grid size  & success probability  & Steps & success probability  & Steps \\
\hline
100          & 0.129953 & 55   & 
0.874064 & 37       \\ 
400        & 0.611614    & 59  & 0.963963 & 84     \\ 
900       &   0.601515  &  451  & 0.984277 & 128        \\ 
1600        &   0.592153  & 617   & 0.986564 & 183      \\ 
2500      &   0.584134 &  1707   & 0.986460 & 229       \\ 
3600       &    0.570086 &  571   & 0.992471 & 273      \\ 
4900  &  0.569388    & 655        & 0.993541 & 319    \\ 
6400 &  0.563390  &  761         & 0.991610 & 368     \\ 
8100       &     0.557533 & 857   & 0.994785 & 413     \\ 
10000      &      0.553369 & 949  & 0.997134 & 471  \\     
\hline
\end{tabular}
\label{Proposed_vs_Grover_k=9}}
\end{table}

\begin{figure}[ht!]
    \centering
    \includegraphics[scale=0.6]{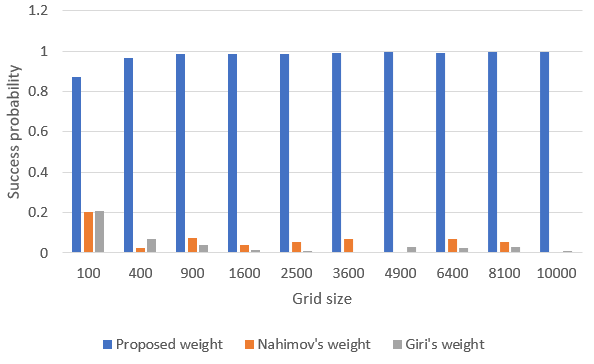}
    \caption{Comparative analysis of success probability  of the proposed LQW for $k = 9$ and different values of grid size $N$ with (i) the proposed $l=\dfrac{4}{N*(k+1)} \pm \delta$,  (ii) $l=\dfrac{4(k-\sqrt{k})}{N}$ \cite{nahimovs2018lackadaisical}, and (iii) $l=\dfrac{4*k}{N}$ \cite{Giri_2019}.}
    \label{fig:compall}
\end{figure}

We have compared our work with \cite{nahimovs2018lackadaisical, Giri_2019} and the results are shown in Table \ref{tbl:nahi_vs_giri_vs_propose_k=9}. In \cite{nahimovs2018lackadaisical, Giri_2019}, the authors tried to generalize the lackadaisical quantum walk approach for multiple marked states by suggesting the optimal weight for self-loop with $l=\dfrac{4(k-\sqrt{k})}{N}$ \cite{nahimovs2018lackadaisical} and $l=\dfrac{4*k}{N}$ \cite{Giri_2019}. But, their weights fail miserably to find a single marked state of this exceptionally configured clustered marked states. In Figure \ref{fig:compall}, we show a comparative analysis of our proposed approach versus \cite{nahimovs2018lackadaisical, Giri_2019} with respect to the success probability for $k = 9$ in different grid sizes from $10 \times 10$ to $100 \times 100$.

\begin{table}[ht!]
\caption{Success probability and Number of steps taken by the proposed LQW for $k = 9$ and different values of grid size $N$ with (i) $l=\dfrac{4(k-\sqrt{k})}{N}$ \cite{nahimovs2018lackadaisical}, (ii) $l=\dfrac{4*k}{N}$ \cite{Giri_2019}, and (iii) the proposed $l=\dfrac{4}{N*(k+1)} \pm \delta$.}
\begin{adjustbox}{max width=\textwidth}
\begin{tabular}{@{}|c|ccc|ccc|ccc|@{}} \hline
Grid size & Weight \cite{nahimovs2018lackadaisical} & success probability  & Steps & Weight \cite{Giri_2019} & success probability  & Steps & Proposed Weight & success probability  & Steps  \\
\hline
100     & 0.24    &       0.203058 & 74 & 0.36 & 0.207837    & 21  & 0.0044    & 0.874064 & 37       \\ 
400       & 0.060000   & 0.025015    & 31  &  0.090000    & 0.066362     & 41 & 0.001    & 0.963963 & 84      \\ 
900       & 0.026667   & 0.076135    & 130  & 0.04    & 0.041122    & 61 & 0.000490    & 0.984277 & 128    \\ 
1600       & 0.015   & 0.041122    & 313 & 0.0225    & 0.015803     & 144 & 0.00025   & 0.986564 & 183      \\ 
2500       & 0.0096   & 0.053291    & 393  & 0.0144    & 0.010333     & 135 & 0.00015    & 0.986460 & 229     \\ 
3600       & 0.006667  & 0.069063    & 267  & 0.01    &  0.003479     & 291& 0.000111   & 0.992471 & 273    \\ 
4900       & 0.004898  & 0.002277    & 253 & 0.007347    & 0.028283    & 147& 0.000082    & 0.993541 & 319     \\ 
6400       & 0.00375  &  0.066896    & 787  & 0.005625    & 0.025452     & 337 & 0.000060    & 0.991610 & 368     \\ 
8100       & 0.002963  &  0.052395    & 727  & 0.00444    & 0.027776     & 191 & 0.000049   & 0.994785 & 413  \\ 
10000      & 0.0024  & 0.000764   & 2149 & 0.0036    & 0.011434     & 361 & 0.000043    & 0.997134 & 471     \\ 
\hline
\end{tabular}
\end{adjustbox}
\label{tbl:nahi_vs_giri_vs_propose_k=9}
\end{table}

Further, we have simulated the result for $k = 25, 49$ (i.e., $25$ marked states arranged in a $5 \times 5$ and $49$ marked states arranged in a $7 \times 7$ cluster), using the same weight of the self loop $l=\dfrac{4}{N*(k+1)} \pm \delta$, where $\delta$ varies between $-0.000004$ to $0.000016$ for different grid sizes, which are given in Table \ref{tbl:proposed_k=25&49}. Hence, from Table \ref{tbl:k=9} and Table \ref{tbl:proposed_k=25&49}, by analysing the value of $\delta$, we infer that the weight of the self loop, $l=\dfrac{4}{N*(k+1)} \pm \delta$, such that $\dfrac{4}{N*(k+2)} \le l \le \dfrac{4}{N*k}$. Thus, we can conclude that our algorithm works for any value of $k$, where $k=n^{2}$, and $n$ an odd integer. Grover's coin based quantum walk \cite{nahimovs2015exceptional} or existing lackadaisical quantum walk \cite{nahimovs2018lackadaisical, Giri_2019} approaches fail to find any of the marked states if $k>9$ for this exceptional configuration. But, according to the trend shown in Table \ref{tbl:proposed_k=25&49}, although our algorithm can achieve a success probability $\sim 1$, the number of steps for marked states $k>25$ grows significantly. We plan to explore the reason for such an increase beyond $k > 25$, and propose a suitable loop weight to reduce the number of steps.

This proposed work has been simulated on Python 3.7, with processor Intel(R) Core(TM) i5-6300U CPU \@ 2.40 GHz 2.50 GHz, RAM 8.00 GB, and 64-bit windows operating system. For higher grid size with more marked states will require higher configuration of the system with respect to both processor and memory.

\begin{table}[ht!]
\centering
\caption{Success probability  and the number of steps taken by the proposed LQW with $l=\dfrac{4}{N*(k+1)} \pm \delta$ for $k = 25$ and $k=49$ in different values of grid size $N$.}
{\begin{tabular}{@{}ccccccc@{}} \hline
Marked-states size & Grid size & $\dfrac{4}{N*(k+1)}$ & $\delta$ & Weight ($l$) & success probability  & Steps  \\
\hline
          
&400   & 0.000384 & 0.000016   & 0.000400   & 0.946912    & 1135          \\ 
$\sqrt{25} \times \sqrt{25}$&900      & 0.000170 &  -0.000004  & 0.000166  & 0.894135    & 4529      \\ 
&1600  & 0.000096 &    0.000001    & 0.000097     &    0.935753    & 6009     \\ 
&2500    & 0.000061 &   0.000001   & 0.000062   & 0.991250  & 18557   \\ 
\hline
$\sqrt{49} \times \sqrt{49}$ &400     &  0.000200 &  0.000004   & 0.000204   &  0.896939  & 234022        \\ 
\hline
\end{tabular}
\label{tbl:proposed_k=25&49}}
\end{table}

\section{Conclusion}

In this paper we have studied the application of quantum random walk on an $\sqrt{N} \times \sqrt{N}$ grid, where the marked states are arranged in a $\sqrt{k} \times \sqrt{k}$ cluster and $k=n^{2}$ for all values of $n \ge 1$. Our results are completely based on numerical simulations which show that using lackadaisical quantum walk, where the weight of the self loop ranges between $l=\dfrac{4}{N*(k+1)} \pm \delta$, where $\dfrac{4}{N*(k+2)} \le l \le \dfrac{4}{N*k}$ (correct up to 6 decimal places), the total success probability of finding all marked states becomes close to 1 in time less than that of quantum walk with no self loop for any odd value of $n \ge 1$, where the marked states are arranged in a $\sqrt{k} \times \sqrt{k}$ cluster and $k=n^{2}$. Furthermore, we have showed by simulation that the weights of \cite{nahimovs2018lackadaisical, Giri_2019} fail to find any of the mark states in this exceptional configuration. Lastly, we have showed that our proposed range of weight is also optimal for higher values of $k$, whereas Grover's coin fails to find any of the marked states when $k > 9$.

\section*{Acknowledgement}
We gratefully acknowledge fruitful discussion regarding quantum walk with Prof. Andris Ambainis, University of Latvia. We also thankful to Dr. Alexander Rivosh, Dr. Nikolajs Nahimovs, and Dr. Raqueline Azevedo Medeiros Santos for the discussion on exceptional configurations of quantum walks.

\bibliographystyle{unsrt}  
\bibliography{references}  

\end{document}